# High-Yield, Scaling-Up Fabrication of Fermi-Level-Pinning-Free Organic Thin-Film Transistor Arrays with Printed Van der Waals Contacts

Zhiyun Wu[1, 2], Shuiren Liu[1, 2], Juzhong Zhang[1], Hanyu Jia[1], Qingqing Sun[1], Xiaoguang Hu[1], Lingxian Meng[1], Xuying Liu[1, *]

[1]School of Materials Science and Engineering, Zhengzhou Key Laboratory of Flexible Electronic Materials and Thin-Film Technologies, Zhengzhou University, Zhengzhou, 450001 China.

[2]These authors contributed equally: Zhiyun Wu, Shuiren Liu

*Corresponding author. E-mail: liuxy@zzu.edu.cn.

Fermi-level pinning (FLP) effect was widely observed in thin-film transistors (TFTs) with van der Waals (vdW) layered semiconductors (organic or two-dimensional) when contact electrodes were thermally evaporated[1-3]. Intensive investigation was implemented for formation of FLP-free interfacial states by eliminating chemical disorder and crystal defects arising from metal deposition[4-9]. However, technical and principal challenges are still existing towards high-yield, wafer-scalable and low-cost integration of TFT devices. Herein, we developed a general, scaling-up strategy to fabricate large-scale, high-performance FLP-free organic TFT (OTFT) arrays by using printed vdW contacts consisting of MXene composite electrodes and 2, 7-dioctyl [1] benzothieno [3, 2-b] [1] benzothiophene (C8BTBT). Room-temperature processes allow for a physically stacked junction without any structural or chemical damages. The OTFT arrays can be printed on a large-area silicon wafer or plastic film with 100% yield, exhibit ultrahigh field-effect mobility ($\mu_{FE}$) over 17.0 square centimetres per volt per second (cm$^2$ V$^{-1}$s$^{-1}$), high on/off ratio exceeding 10$^8$, relatively low contact resistance of 3k ohm micrometres. The underlying mechanism for the high device performance was unveiled by Kelvin Probe Force Microscopy (KPFM) combined with theoretical simulation. The results indicate that work function (W$_F$) of the printed electrodes can be tuned at a wide range of 4.8-5.6 eV, thus significantly lowering the charge-injection barrier at the contact interfaces with ideal FLP-free character (the interfacial factor reaches 0.99 ± 0.02). This study paves a general strategy for achieving large-scale, high-performance thin-film electronics.

Organic semiconductors with low-temperature processability show intriguing applications in low-cost, scaling-up fabrication of electronic devices. In past decades, in order to unveil the fundamental charge transporting principles of semiconductor-metal interface in high-performance electronic devices, the Schottky barrier height (SBH, $\phi_B$) for charge injection from the metal electrode to the semiconductor has often been extracted as one of the most crucial electrical characteristics[1,3,5]. However, in organic electronics, the metallization within the first few molecular layers of organic semiconducting thin films during electrode deposition was confirmed to induce gap states, interfacial chemical and physical defects[10]. As a consequence, an interface dipole ($P$) can be spontaneously formed between the organic semiconductor and the metal surface; thus, the Fermi level ($E_F$) was always found to be pinned, so that the theoretically calculated $\phi_B$ values are not consistent with the experimental observation[11].

To date, a wide variety of interfacial tuning methods for contact-semiconductor systems have ever been reported to eliminate the FLP, including introduction of a doping interlayer[12,13] or formation of vdW contact via transferring metal electrode[5-7]. Both of them are intensively implemented in organic electronics to inhibit metal clusters to penetrate into the semiconductor thin film, and are confirmed to allow for offering a clean, damage-free metal/semiconductor interface to significantly reduce the $\phi_B$ values. However, an ideal metal/semiconductor contact with negligible FLP effect has not yet been realized experimentally, since there is still a lack of metal deposition strategies to minimize the energy offset between the $E_F$ and the onset of the lowest unoccupied molecular orbital (LUMO)/the highest occupied molecular orbital (HOMO) in organic semiconductors.

In this study, we report scaling-up fabrication of wafer-scale, high-performance, FLP-free OTFT arrays by using screen-printed vdW contacts with a physically stacked metal/semiconductor junction. A printable conductive ink (M-PBr) consisting of water base, MXene sheets and the polymer dopant (Poly [(9,9-bis (3'-((N, N-dimethyl)-N-ethylammonium)-propyl)-2,7-fluorene)-alt-2,7-(9,9-dioctylfluorene)] dibromide, PFN-Br, 0.1-2 wt%) was prepared as illustrated in Fig. 1a. The lateral size and thickness of the used MXene nanosheets were approximately 3 μm and 2 nm . Comprehensive analysis based one X-Ray diffraction (XRD), Raman spectra and X-ray photoelectron spectroscopy (XPS) proved the efficient etching from $Ti_3AlC_2T_x$ MAX to $Ti_3C_2T_x$ MXene and the doping by PFN-Br as well. Moreover, the introduction of dopants does not change the lattice parameters and the in-plane and out-of-plane vibration peaks of titanium atoms of M-PBr ink compared to MXene ink. In particular, the PFN-Br acts as charge carrier transporting material for high-performance photovoltaic devices[14], which bears with ionic pendants enabling to adjust the surface tension and the rheological property of the aqueous ink . By using a screening printer (Fig. 1b), high-precision (30-200 mm) conductive electrodes can be deposited on various substrate surfaces with a wide range of surface energy (10-1000 $mJ/m^2$), including silicon dioxide ($SiO_2$), coated paper and polyethylene terephthalate (PET) .

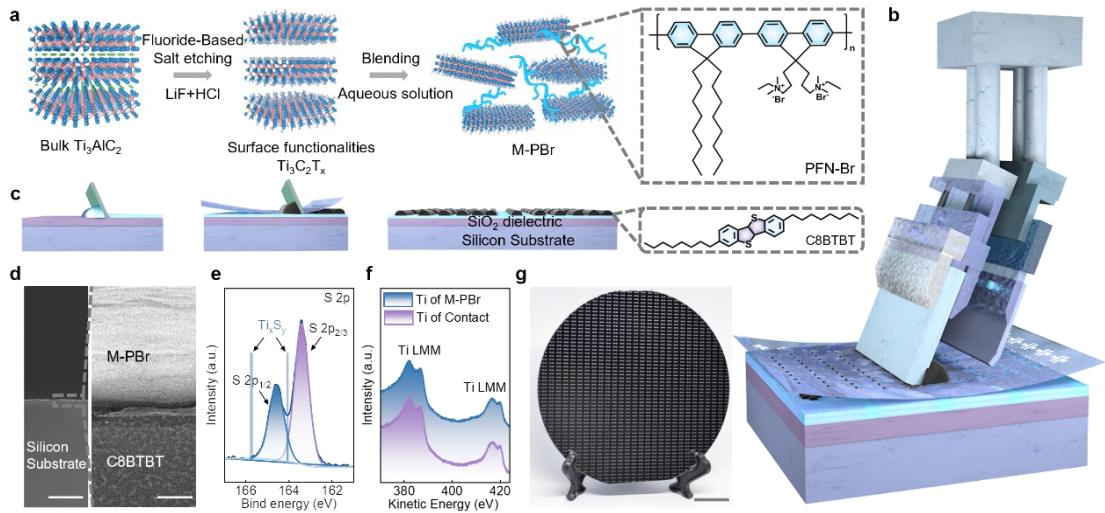

**Fig 1. Large-area, high-yield fabrication of OTFT arrays with printed vdW contacts. a,** Schematic of the synthesis of MXene from the MAX phase by in-situ etching and the following doping process with PFN-Br. **b,** Schematic of the deposition and the formation of M-PBr thin films by screen printing. **c,** Schematic illustration of device fabrication, including solution method for semiconductor growth, printing method for electrode deposition. **d,** Cross-sectional SEM and TEM images of the screen-printed metal-semiconductor contact. Scale bars, 30 μm and 5 nm, respectively. **e,** XPS of the S element at the M-PBr/C8BTBT contact. **f,** X-ray induced Auger spectrum of the Ti elements at M-PBr and the M-PBr/C8BTBT contact. **g,** Photograph of an integrated M-PBr electrode array on a 4-inch silicon wafer. Scale bar, 3 mm.

When depositing M-PBr electrodes on the surface of C8BTBT thin film to fabricate bottom-gate top-contact OTFT devices (Fig. 1c), all processes can be carried out at room temperature, which guarantees the damage-free, chemical-bonding-free fabrication of vdW contacts with electrostatic interaction. Scanning Electron Microscopy (SEM) and Transmission Electron Microscopy (TEM) showing the cross-sectional profiles of the resulting vdW interface are displayed in Fig. 1d, which provide with obvious evidences for the formation of a clean contact. Further chemical analysis shows that the interfacial S element in X-ray photoelectron spectroscopy (XPS) retains its original characteristic peak, and there is no Ti-S bond (Fig. 1e). Ti LMM Auger measurements at the M-PBr/C8BTBT interface were determined to be similar to the one in M-PBr, suggesting that the electronic structure of Ti at the contact point was not perturbed (Fig. 1f). It is worth noting that direct printing of conductive patterns on organic semiconducting films was rarely reported due to the challenging issue in the controllable capturing, fracturing and coalescing of the ink microdroplets. Therefore, many studies preferred to adopt bottom-contact configuration; however, the discrepancy in the surface energy of metal electrodes and the substrates normally interrupts the continuous solution-processing crystallization of organic semiconductors to involve the OTFT in remarkable enhanced FLP effect. This result has been interpreted that the disordered interfacial molecular packing extended the tails of Density of States (DOS) in HOMO and LUMO into the energy gap, which is responsible for an induced gap state.

In comparison to the transferring method that has been widely used in two-dimensional vdW devices, the screen-printing strategy associated with the aqueous M-PBr ink allows for scaling-up patterning of conductive electrodes on the top surfaces of various organic semiconducting films to fabricate large-area OTFT arrays. Wafer-scale, ultra-thin C8BTBT films prepared by blade shearing were selected to demonstrate the rapid operation speed and the high uniformity of the printed top-contact electrodes. One of the OTFT arrays in 4-inch wafer size (the maximum size limited by our screening printer) containing more than $50 \times 50$ M-PBr-vdW source-drain electrode pairs (Fig. 1g) was printed out with the speed of 2 inches per second. Image similarity analysis shows the M-PBr electrode array displays a narrower reproducibility range of 90-105% than the original MXene . Such a high compatibility with the different surfaces was investigated to primarily attribute to a three-dimensional interconnected network, which sustains the initial shape of the printed patterns, and suppresses the occurrence of nanosheet slipping in MXene composites[15]. Furthermore, in the simulated printing shearing experiment, the doped ink with higher viscosity shows typical non-Newtonian fluid behavior; the dopant PFN-Br is able to reduce the surface energy difference between water-based ink and organic semiconductor crystal, which is beneficial for rapidly forming the well-defined patterns of M-PBr.

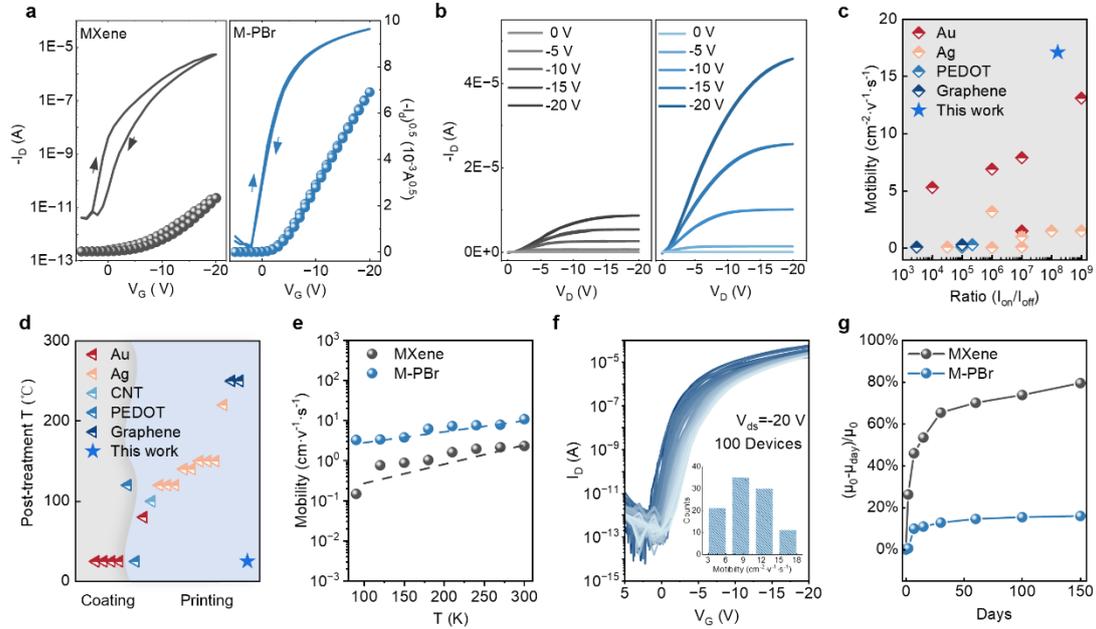

**Fig 2. Electrical performance of the printed vdW-OTFT devices. a,** Transfer characteristics of MXene and M-PBr OFETs at $V_D$= -20V, where the channel length (width) is 100 μm (600 μm). The calculated results show the field-effect mobility of M-PBr OTFT up to 17.1 cm$^2$ V$^{-1}$ s$^{-1}$. **b,** Output characteristics of MXene and M-PBr OFETs. **c, d** Comparison of mobility, on/off ratio **(c)**[16-31], electrode processing temperature **(d)**[16-22,25,27-31] of M-PBr OFETs with values reported in the literature using different methods. **e,** Charge-carrier mobility as a function of temperature. **f,** Transfer characteristics of 100 M-PBr OTFTs. The inset is the histogram of field-effect mobility obtained from M-PBr OTFTs. **g,** Attenuation rate of MXene and M-PFN OTFTs over time.

The field-effect performance of the obtained OTFT arrays was characterized to quantitatively evaluate the charge carrier injection ability in the printed vdW contact. When the top source-drain electrode was doped by PFN-Br, on-state current of the OTFT remarkably increased by one order of magnitude comparing with non-doped devices (Fig. 2a), thereby leading to the highest $\mu_{FE}$ value of 17.1 cm$^2$ V$^{-1}$s$^{-1}$ and high on/off ratio exceeding 10$^8$. Besides, the doped contacts obviously suppressed the hysteresis (Fig. 2a), resulted in a steep subthreshold region with a slop of less than 200 mV/ Dec; the I–V curves tend to be linear at small source–drain voltages towards a typical Ohmic behavior (Fig. 2b). In particular, the obtained $\mu_{FE}$ reaches the record value in all OTFT devices with printed contacts (Fig. 2c); moreover, all processes can be carried out at the most ambient atmosphere without any post-treatment, while most conductive inks need to experience high-temperature or laser sintering for forming high conduction and steady adhesion (Fig. 2d).

We further explore the transmission types of the devices by analyzing the temperature dependence of the mobility. Arrhenius fitting indicates that the mobility is positively correlated with temperature, indicating that the printed vdW OTFT has an activated behavior (Fig. 2e). The activated thermal behavior is generally assigned to the electronic state existing in each layer of the device, which is in accordance with the typical charge transport mechanism of organic semiconductors[13]. We deduce activation energy ($E_a$) according to the Nernst−Einstein equation:

$$\sigma(T) = \frac{\sigma_0}{T} \exp\left(-\frac{E_a}{K_B T}\right) \quad (1)$$

in which $\sigma, T\ and\ K_B$ represent the conductivity, the absolute temperature (K), and Boltzmann's constant (known as physical constant equal to 8.617 × 10$^{-5}$ eV/°C). The device with M-PBr has a smaller activation energy of 0.06 eV than that of the one with MXene (0.14 eV), indicating that the energy required for the carrier activation transition in the M-PBr/C8BTBT contact is relatively low, which leads to a high carrier transfer efficiency. The curves of device mobility versus drain voltage ($V_D$) follow the power law, where M-PBr has a lower power law index than MXene, implying enhanced charge carrier transport performance.

The uniformity and stability of electrical properties in the printed vdW- OTFT can be demonstrated by investigating the $\mu_{FE}$ distribution and the relative attenuation rate over time, respectively. The narrow $\mu_{FE}$ distribution ($\bar{\mu}_{FE}$=10.2 cm$^2$ V$^{-1}$s$^{-1}$, Confidence interval is (9.5, 10.9)) of 100 M-PBr OFETs displayed in Fig. 2f suggests that a good device uniformity, which can be comparable with the reported OTFT arrays[27]; When the printed OTFT arrays were stored in air for 5 months, the M-PBr TFT exhibited a less than 10% mobility decay (nearly 80% for the MXene OTFTs) (Fig. 2g). The high electrical stability can be understood by correlating the conductivity with the varying temperature to deduce activation energy ($E_a$). A significant increase (from 0.016 eV to

0.028 eV) in the $E_a$ value was extracted for the M-PBr doped contacts, indicating a higher electrical stability than the MXene ones, which is consistent with the observation (Fig. 2g).

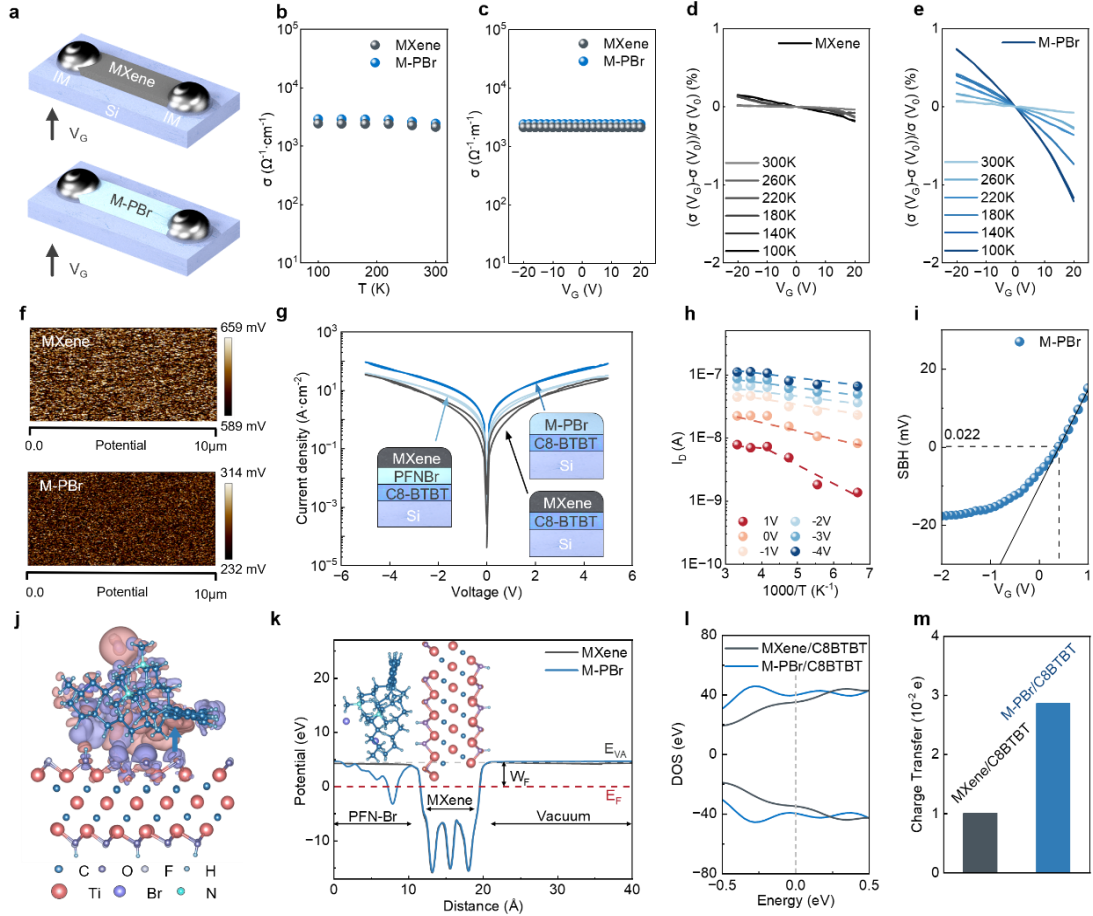

**Fig 3. The operation mechanism of the printed vdW contacts for high-performance OTFT arrays. a,** The sketch of screen-printed MXene and M-PBr devices. **b, c,** Conductivity σ at gate voltage $V_G$ = 0V as a function of temperature T **(b)** and conductivity σ as a function of the gate voltage $V_G$ **(c)** for printed MXene and M-PBr device. **c,** Conductivity ratio as a function of $V_G$ upon variable temperature in the printed MXene ink device **(d)** and printed MXene ink device **(e)**. **f,** KPFM images of pristine MXene electrode and M-PBr electrode. Obtained results show the function is 5.12eV and 5.46eV. **g,** Current density curves of Si/SiO$_2$/Spin PFNBr/MXene, Si/SiO$_2$/MXene, Si/SiO$_2$/M-PBr devices. (The electrode area is 0.12 mm$^2$.) **h,** Arrhenius-type plot constructed using part a at different gate voltages. **i,** Extracted effective barrier height of M-PBr/C8BTBT contact as a function of $V_G$. **j,** The differential charge density of MXene and PFN-Br contacts. The direction of the dipole moment is from MXene to PFN-Br, with the electrons gaining in the purple color and the electrons losing in the pink color. **k,** Electrostatic potential averaged over planes perpendicular to the MXene and M-PBr interface. The computed structures are shown within the plots, The gray dashed lines represent the dipole corrected vacuum levels. The Fermi energy is set to zero, so that the vacuum potential, just away from the MXene/M-PBr surface, corresponds to the $W_F$, of the MXene/M-PBr surface, as depicted in the panels. **l,** Density of states (DOS) for MXene/C8BTBT and M-PBr/C8BTBT. The Fermi level is set at zero. **m,** The charge transfer from MXene/M-PBr to C8BTBT by Bader charge analysis.

To gain an unambiguous understanding for operation mechanism underlying the enhanced field-effect characters, electrical properties of the printed vdW contacts are systematically studied. The gate modulated electrical conductivity of MXene and M-PBr was determined with variable temperature by using a device configuration with liquid metal as touch panel (Fig. 3a). As a result, both two channels exhibited slightly temperature-independent and gate-bias-independent conduction manners, indicating a metallic feature with high thermal stability (Fig. 3b, c). Therefore, doping of low-concentration PFN-Br was found to hardly tune the intrinsic metallic conductive property of MXene nanosheets. On the other hand, conductivity variation ratio extracted from $[\sigma(V_G) - \sigma(V_0)]/\sigma(V_0)$ presents a remarkable difference upon changing $V_G$ and T after doping of PFN-Br (Fig. 3d, e shows the full set of σ versus $V_G$ curves). This field-effect current observed in the M-PBr device indicates that organic semiconducting dopant can act as linking bridge to ensure charge carrier to transport across the gaps between the adjacent MXene nanosheets. The $V_G$-dependent conductivity ratio decreases monotonically with increasing $V_G$, indicating remarkable hole conduction induced by the doping of adsorbed molecules. Nearly negligible hysteresis suggests charge transfer with traps is quiet few in the printed film, primarily concentrated at the M-PBr/SiO$_2$ interface.

The enhanced charge injection capability in the printed vdW contacts was believed to be attributed to the tuned energy barrier at the electrode/ semiconductor interfaces. In an ideal device, the barrier can be theoretically predicted by the Schottky–Mott rule, $\phi_B = I_s - W_F$, $I_s$ and $W_F$ are the ionization potential of semiconductors and the work function of the metal, respectively[15]; therefore, the $W_F$ of printed electrodes needs to be close to the HOMO level (-5.49 eV) of C8BTBT in the OTFT devices, thus minimizing the value of $\phi_B$. As determined by KPFM (Fig. 3f), the MXene shows W$_F$ around -5.12 eV; this value decreased down to a range from -5.20 to -5.53 eV for M-PBr, corresponding to the doping rate. As widely reported, the $W_F$ levels of MXene films are sensitive to the surroundings, which are able to induce surface polarization[15]. M-PBr electrode with the W$_F$ of -5.41 eV was employed to construct a sandwiched device with three different configurations of MXene/C8BTBT/MXene, MXene/PFN-Br/C8BTBT/PFN-Br/MXene, and M-PBr/C8BTBT/M-PBr. According to the expression: $\Delta\emptyset_B = -(e/\varepsilon_0)\Delta P$, where $\emptyset_B$, $e$, $\varepsilon_0$, and $P$ represent the W$_F$, the amount of electron charge, the vacuum dielectric constant and the surface dipole moment, respectively; thus, the change of surface dipole moment, $\Delta P$, was calculated to be $4.11 \times 10^{-4}$ $(e/Å)$, indicating a weakly induced interfacial state. Note that the current-voltage characteristics show that the doped electrode of M-PBr is capable of yielding a significant decrease in contact resistance than the others (Fig. 3g), mainly ascribed to the minimized energy barrier and interfacial dipole moment.

To determine true $\emptyset_B$ values, temperature-dependent current measurement was carried out on the OTFT devices with M-PBr contacts (Fig. 3h). At a low gate voltage (V$_G$), the current was obviously relaying on the temperature, suggesting thermionic emission (TE) as the dominant mechanism; while at a high V$_G$ regime, field emission (FE) dominates and current becomes independent of temperature as the channel gets electrically doped due to large back gate voltages. Arrhenius plots for various V$_G$ in the TE and FE regime are shown in Fig. 3i. Using standard TE theory[32], the Schottky Barrier Height value can be extracted based on the following equation:

$$I_{ds} = AA^* \times T^2 \exp\left[-\frac{\emptyset_{SB}}{kT}\right] \quad (2)$$

where $I_{ds}$ is the source-drain current. $A$ is the junction area. $A^*$ is the Richardson constant. $K$ is the Boltzmann constant. $T$ is the temperature. We noted that all the SBHs extracted using this formula are determined from the flat-band condition. SBH extraction value of M-PBr OTFTs is 0.022 eV, which is located at the same order of magnitude with the theoretical calculation (0.08 eV). Such a low SBH was believed to introduce low contact resistance, and thus to lead to the linear behavior in IV curves in Fig. 2b.

Theoretical investigation for revealing the mechanism of tuning W$_F$ of M-PBr was implemented via Density functional theory (DFT) calculation. Charge density difference plots associated with the planar averaged charge density (Fig. 3j) visualize interfacial charge transfer from the PFN-Br molecule to the MXene nanosheet, the charge transfer amount is 0.11e . As a result, redistribution of electrostatic interfacial potentials was induced accordingly to generate a dynamic equilibrium dipole moment (Fig. 3k), thus leading to tunable $W_F$. When contacting with C8BTBT thin films, the total DOS of the system with M-PBr (39.64 eV) was observed to be higher than that of the one with MXene (34.75 eV) (Fig. 3l); moreover, the simulated partial density of states indicates that the DOS values of all elements in the former are much higher than that in the later . In particular, the interfacial charge transfer was enhanced to be three times higher after the electrodes were doped by PFN-Br (Fig. 3m), which was quantitatively analyzed according to charge balance principle[33].

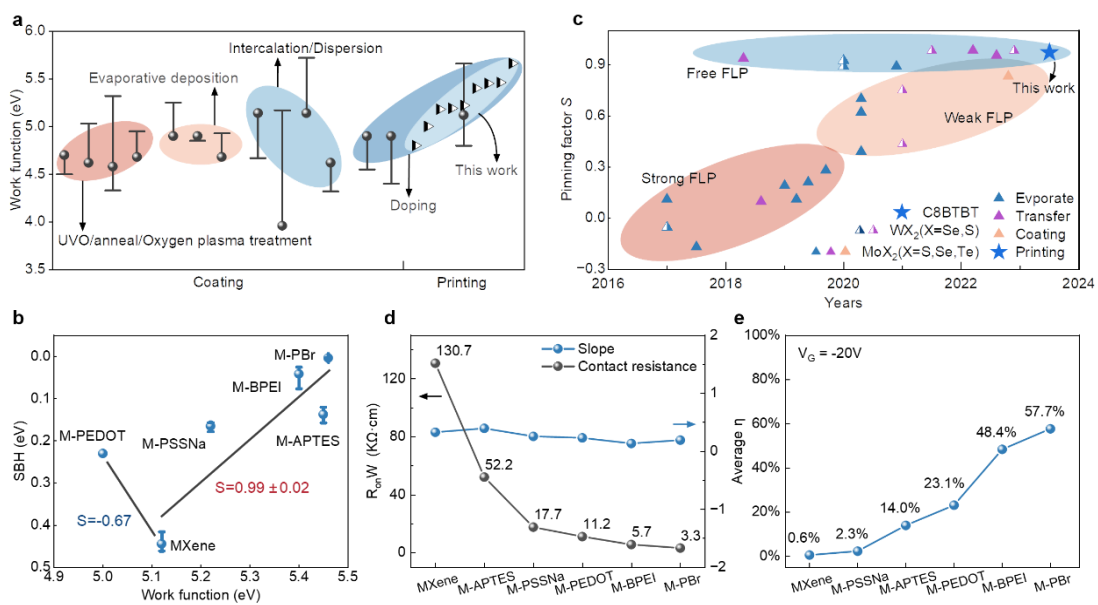

**Figure 4. Generalization of the proposed strategy to a wide variety of printable vdW contacts. a.** $Ti_3C_2T_x$ $W_F$ regulation values and approaches found in the literature compared with those obtained in this work[34-43]. **b,** $\varnothing_B$ values experimentally determined for printed electrodes. **c,** Summary of FLP and contact parameters of metal-vdW semiconductors in the recent five years[5,39,44-50]. **d,** Contact resistance of different printed OTFT devices. **e,** Injection efficiency of OTFT devices with the printed doped-MXene contact electrodes.

This strategy for tuning the $W_F$ values in MXene-based electrodes is feasible and reliable when various dopants, including branched polyethyleneimine (BPEI), PSS:PEDOT, and poly(sodium-4-styrenesulfonate) (PSSNa), have been employed. The introduction of characteristic groups of dopants and organic molecular chains also optimizes the rheological properties and surface energy of the ink. It is worth noting that the basic principle of doping is to manipulate the surface potential of the pristine MXene while maintaining its interlayer structure. XRD and Raman spectra show that lattice and atomic vibrations have not changed, proving that our doping can maximize the intrinsic charge transport of MXene. By stabilizing the surface groups using different dopants, all of them show higher activation stability than MXene. As a reference, 3-Aminopropyltriethoxysilane (APTES) as surface modifier was also used to highlight the generalization of printed vdW contacts in high-performance organic electronic devices. The $W_F$ values of the printed doped-MXene electrodes cover a wide range from -4.60 to -5.65 eV, which can be comparable to the reported values (Fig. 4a) in the printed vdW OTFT devices. When these doped-MXene contacts were utilized in printed OTFTs, the true SBHs can be extracted as described in Fig. 4b, which are strongly dependent on the metal $W_F$ with a slope ($S = 0.99\pm0.02$) approaching the unity, suggesting ideal obedience to the Schottky–Mott law. Only the one with PSS:PEDOT (M-PEDOT) exhibited the FLP effect with $S = -0.67$, which was speculated to originate from a potential chemical bonding between PSS chains with BTBT cores. This hypothesis was supported by the result of PSSNa, which is believed to physically interact with the organic semiconducting molecules to form FLP-free OTFTs. Further understanding for this phenomenon is still under investigation. Despite the fact that the free-FLP TFTs can be often achieved with 2D materials, the printed vdW OTFT with doped MXene electrodes was proposed for the first time (Fig. 4c).

Both the extracted contact resistance and charge injection efficiency exhibited remarkable SBH dependence (Fig. 4d, e). To reduce the Schottky barrier with interfacial engineering was powerful way to achieve low contact resistance, thus yielding a high injection efficiency. In this study, the M-PBr vdW OTFTs show a low contact resistance of 3.3 K (Fig. 4d) with high injection efficiency exceeding 60% (Fig. 4e). These values can be comparable to the reported benchmark devices. The obtained contact resistance is a low value in OTFT devices, but is still higher than the devices with 2D organic semiconducting films mainly due to high bulk electrical defects. Therefore, the electrical performance of printed vdW OTFTs with doped MXene contacts will be further improved by utilization of organic semiconductor crystal growth engineering. Furthermore, other semiconductors, including 2D materials, carbon nanotubes, and metal oxides, can be selected as the active layers for fabricating high-performance TFT devices by utilization of the proposed strategy.

References:

1	Wang, Y. *et al.* P-type electrical contacts for 2D transition-metal dichalcogenides. *Nature* **610**, 61-66, doi:10.1038/s41586-022-05134-w (2022).

2	Shen, P. C. *et al.* Ultralow contact resistance between semimetal and monolayer semiconductors. *Nature* **593**, 211-217, doi:10.1038/s41586-021-03472-9 (2021).

3	Wang, Y. *et al.* Van der Waals contacts between three-dimensional metals and two-dimensional semiconductors. *Nature* **568**, 70-74, doi:10.1038/s41586-019-1052-3 (2019).

4	Chen, P. *et al.* Approaching the intrinsic exciton physics limit in two-dimensional semiconductor diodes. *Nature* **599**, 404-410, doi:10.1038/s41586-021-03949-7 (2021).

5	Liu, Y. *et al.* Approaching the Schottky-Mott limit in van der Waals metal-semiconductor junctions. *Nature* **557**, 696-700, doi:10.1038/s41586-018-0129-8 (2018).

6	Liu, L. *et al.* Transferred van der Waals metal electrodes for sub-1-nm MoS2 vertical transistors. *Nature Electronics* **4**, 342-347, doi:10.1038/s41928-021-00566-0 (2021).

7	Jung, Y. *et al.* Transferred via contacts as a platform for ideal two-dimensional transistors. *Nature Electronics* **2**, 187-194, doi:10.1038/s41928-019-0245-y (2019).

8	Wu, F. *et al.* Vertical MoS(2) transistors with sub-1-nm gate lengths. *Nature* **603**, 259-264, doi:10.1038/s41586-021-04323-3 (2022).

9	Song, S. *et al.* Wafer-scale production of patterned transition metal ditelluride layers for two-dimensional metal–semiconductor contacts at the Schottky–Mott limit. *Nature Electronics* **3**, 207-215, doi:10.1038/s41928-020-0396-x (2020).

10	Kahn, Y. H. a. A. Chemistry and electronic properties of metal-organic semiconductor interfaces: Al, Ti, In, Sn, Ag, and Au on PTCDA. *Physical Review B* **54**, 13784 (1996).

11	S. Yogev, R. M., M. Nakamura, U. Zschieschang, H. Klauk, and Y. Rosenwaks. Fermi Level Pinning by Gap States in Organic Semiconductors. *Physical Review Letters* **110**, 036803 (2013).

12	Chuan Liu, G. L., Riccardo Di Pietro, Jie Huang, Yong-Young Noh,Xuying Liu,and Takeo Minari. Device Physics of Contact Issues for the Overestimation and Underestimation of Carrier Mobility in Field-Effect Transistors. *Physical Review Applied* **8**, 034020 (2017).

13	Chuan Liu, K. H., Won-Tae Park, Minmin Li, Tengzhou Yang, & Xuying Liu, L. L., Takeo Minaric and Yong-Young Noh. A unified understanding of charge transport in organic semiconductors: the importance of attenuated delocalization for the carriers. *Materials Horizons* **4**, 608 (2017).

14	Haoran Tang, Y. B., Haiyang Zhao, Xudong Qin, Zhicheng Hu, Cheng Zhou, Fei Huang and Yong Cao. Interface Engineering for Highly Efficient Organic Solar Cells. *Advanced Materials*, 2212236 (2023).

15	Zhiyun Wu, S. L., Zijuan Hao, and Xuying Liu. MXene Contact Engineering for Printed Electronics. *Advanced Science*, 2207174 (2023).

16	Cao, C., Andrews, J. B., Kumar, A. & Franklin, A. D. Improving Contact Interfaces in Fully Printed Carbon Nanotube Thin-Film Transistors. *ACS Nano* **10**, 5221-5230, doi:10.1021/acsnano.6b00877 (2016).

17	Kwon, J. *et al.* Three-Dimensional, Inkjet-Printed Organic Transistors and Integrated Circuits with 100% Yield, High Uniformity, and Long-Term Stability. *ACS Nano* **10**, 10324-10330, doi:10.1021/acsnano.6b06041 (2016).

18  Song, D. *et al.* High-Resolution Transfer Printing of Graphene Lines for Fully Printed, Flexible Electronics. *ACS Nano* **11**, 7431-7439, doi:10.1021/acsnano.7b03795 (2017).

19  Hyun, W. J. *et al.* All-Printed, Foldable Organic Thin-Film Transistors on Glassine Paper. *Adv Mater* **27**, 7058-7064, doi:10.1002/adma.201503478 (2015).

20  Liu, X. *et al.* Spontaneous Patterning of High-Resolution Electronics via Parallel Vacuum Ultraviolet. *Adv Mater* **28**, 6568-6573, doi:10.1002/adma.201506151 (2016).

21  Pierre, A. *et al.* All-printed flexible organic transistors enabled by surface tension-guided blade coating. *Adv Mater* **26**, 5722-5727, doi:10.1002/adma.201401520 (2014).

22  Grau, G., Kitsomboonloha, R., Swisher, S. L., Kang, H. & Subramanian, V. Printed Transistors on Paper: Towards Smart Consumer Product Packaging. *Advanced Functional Materials* **24**, 5067-5074, doi:10.1002/adfm.201400129 (2014).

23  Ji, D. *et al.* "Regioselective Deposition" Method to Pattern Silver Electrodes Facilely and Efficiently with High Resolution: Towards All-Solution-Processed, High-Performance, Bottom-Contacted, Flexible, Polymer-Based Electronics. *Advanced Functional Materials* **24**, 3783-3789, doi:10.1002/adfm.201304117 (2014).

24  Makita, T. *et al.* Electroless-Plated Gold Contacts for High-Performance, Low Contact Resistance Organic Thin Film Transistors. *Advanced Functional Materials* **30**, 2003977, doi:10.1002/adfm.202003977 (2020).

25  Minari, T. *et al.* Room-Temperature Printing of Organic Thin-Film Transistors with π-Junction Gold Nanoparticles. *Advanced Functional Materials* **24**, 4886-4892, doi:10.1002/adfm.201400169 (2014).

26  Ji, D. *et al.* Silver mirror reaction for organic electronics: towards high-performance organic field-effect transistors and circuits. *Journal of Materials Chemistry C* **2**, 4142, doi:10.1039/c4tc00119b (2014).

27  Fukuda, K. *et al.* Fully-printed high-performance organic thin-film transistors and circuitry on one-micron-thick polymer films. *Nat Commun* **5**, 4147, doi:10.1038/ncomms5147 (2014).

28  Liu, X. *et al.* Homogeneous dewetting on large-scale microdroplet arrays for solution-processed electronics. *NPG Asia Materials* **9**, 409-416, doi:10.1038/am.2017.123 (2017).

29  Fukuda, K., Takeda, Y., Mizukami, M., Kumaki, D. & Tokito, S. Fully solution-processed flexible organic thin film transistor arrays with high mobility and exceptional uniformity. *Sci Rep* **4**, 3947, doi:10.1038/srep03947 (2014).

30  Shiwaku, R. *et al.* Printed 2 V-operating organic inverter arrays employing a small-molecule/polymer blend. *Sci Rep* **6**, 34723, doi:10.1038/srep34723 (2016).

31  H. Sirringhaus, K., . H. Friend, T. Shimoda, M. Inbasekaran, W. Wu, E. P. Woo. High-Resolution Inkjet Printing of All-Polymer Transistor Circuits. *Science* **290**, 2123-2126 (2000).

32  T., R. Recent advances in Schottky barrier concepts. *Materials Science and Engineering* **35**, 1-138 (2001).

33  Troy Van Voorhis, T. K., Benjamin Kaduk, & Lee-Ping Wang, C.-L. C., and Qin Wu. The Diabatic Picture of Electron Transfer, Reaction Barriers, and Molecular Dynamics. *Annual Reviews* **61**, 149–170 (2010).

34  Wang, J. *et al.* Plasma Oxidized $Ti_3C_2T_x$ MXene as Electron Transport Layer for Efficient Perovskite Solar Cells. *ACS Appl Mater Interfaces* **13**, 32495-32502,

doi:10.1021/acsami.1c07146 (2021).

35  Huang, C., Shi, S. & Yu, H. Work Function Adjustment of Nb2CTx Nanoflakes as Hole and Electron Transport Layers in Organic Solar Cells by Controlling Surface Functional Groups. *ACS Energy Letters* **6**, 3464-3472, doi:10.1021/acsenergylett.1c01656 (2021).

36  Lyu, B. *et al.* Large-Area MXene Electrode Array for Flexible Electronics. *ACS Nano* **13**, 11392-11400, doi:10.1021/acsnano.9b04731 (2019).

37  Wang, H. *et al.* 2D MXene-Molecular Hybrid Additive for High-Performance Ambipolar Polymer Field-Effect Transistors and Logic Gates. *Adv Mater* **33**, 2008215, doi:10.1002/adma.202008215 (2021).

38  Chen, J. *et al.* Work-Function-Tunable MXenes Electrodes to Optimize p-CsCu2I3/n-Ca2Nb3-xTaxO10 Junction Photodetectors for Image Sensing and Logic Electronics. *Advanced Functional Materials* **32**, 2201066, doi:10.1002/adfm.202201066 (2022).

39  Chen, R. S. *et al.* MoS2 Transistor with Weak Fermi Level Pinning via MXene Contacts. *Advanced Functional Materials* **32**, 2204288, doi:10.1002/adfm.202204288 (2022).

40  Schultz, T. *et al.* Surface Termination Dependent Work Function and Electronic Properties of Ti3C2Tx MXene. *Chemistry of Materials* **31**, 6590-6597, doi:10.1021/acs.chemmater.9b00414 (2019).

41  Yu, Z. *et al.* MXenes with tunable work functions and their application as electron- and hole-transport materials in non-fullerene organic solar cells. *Journal of Materials Chemistry A* **7**, 11160-11169, doi:10.1039/c9ta01195a (2019).

42  Schultz, T. *et al.* Work function and energy level alignment tuning at Ti3C2Tx MXene surfaces and interfaces using (metal-)organic donor/acceptor molecules. *Physical Review Materials* **7**, 045002, doi:10.1103/PhysRevMaterials.7.045002 (2023).

43  Zhang, C. *et al.* Large Area and High-Efficiency MXene–Silicon Solar Cells by Organic Enhanced Dispersity and Work Function. *Solar RRL* **6**, 2200743, doi:10.1002/solr.202200743 (2022).

44  Kim, C. *et al.* Fermi Level Pinning at Electrical Metal Contacts of Monolayer Molybdenum Dichalcogenides. *ACS Nano* **11**, 1588-1596, doi:10.1021/acsnano.6b07159 (2017).

45  Kim, G. S. *et al.* Schottky Barrier Height Engineering for Electrical Contacts of Multilayered MoS(2) Transistors with Reduction of Metal-Induced Gap States. *ACS Nano* **12**, 6292-6300, doi:10.1021/acsnano.8b03331 (2018).

46  Jang, J. *et al.* Fermi-Level Pinning-Free WSe(2) Transistors via 2D Van der Waals Metal Contacts and Their Circuits. *Adv Mater* **34**, 2109899, doi:10.1002/adma.202109899 (2022).

47  Yang, Z. *et al.* A Fermi-Level-Pinning-Free 1D Electrical Contact at the Intrinsic 2D MoS(2)-Metal Junction. *Adv Mater* **31**, 1808231, doi:10.1002/adma.201808231 (2019).

48  Murali, K., Dandu, M., Watanabe, K., Taniguchi, T. & Majumdar, K. Accurate Extraction of Schottky Barrier Height and Universality of Fermi Level De-Pinning of van der Waals Contacts. *Advanced Functional Materials* **31**, 2010513, doi:10.1002/adfm.202010513 (2021).

49  Sotthewes, K. *et al.* Universal Fermi-Level Pinning in Transition-Metal Dichalcogenides. *J Phys Chem C Nanomater Interfaces* **123**, 5411-5420, doi:10.1021/acs.jpcc.8b10971 (2019).

50  Wang, Q., Shao, Y., Gong, P. & Shi, X. Metal–2D multilayered semiconductor junctions: layer-number dependent Fermi-level pinning. *Journal of Materials Chemistry C* **8**, 3113-3119, doi:10.1039/c9tc06331e (2020).